# Unleashing the Hidden Power of Compiler Optimization on Binary Code Difference: An Empirical Study

Xiaolei Ren
University of Texas at Arlington, USA
xiaolei.ren@mavs.uta.edu

Michael Ho
University of Texas at Arlington, USA
michael.ho22@mavs.uta.edu

Jiang Ming*
University of Texas at Arlington, USA
jiang.ming@uta.edu

Yu Lei
University of Texas at Arlington, USA
ylei@cse.uta.edu

Li Li
Monash University, Australia
Li.Li@monash.edu

## Abstract

Hunting binary code difference without source code (i.e., binary diffing) has compelling applications in software security. Due to the high variability of binary code, existing solutions have been driven towards measuring semantic similarities from syntactically different code. Since compiler optimization is the most common source contributing to binary code differences in syntax, testing the resilience against the changes caused by different compiler optimization settings has become a standard evaluation step for most binary diffing approaches. For example, 47 top-venue papers in the last 12 years compared different program versions compiled by *default* optimization levels (e.g., -O$x$ in GCC and LLVM). Although many of them claim they are immune to compiler transformations, it is yet unclear about their resistance to *non-default* optimization settings. Especially, we have observed that adversaries explored non-default compiler settings to amplify malware differences.

This paper takes the first step to systematically studying the effectiveness of compiler optimization on binary code differences. We tailor search-based iterative compilation for the auto-tuning of binary code differences. We develop *BinTuner* to search near-optimal optimization sequences that can maximize the amount of binary code differences. We run BinTuner with GCC 10.2 and LLVM 11.0 on SPEC benchmarks (CPU2006 & CPU2017), Coreutils, and OpenSSL. Our experiments show that at the cost of 279 to 1,881 compilation iterations, BinTuner can find custom optimization sequences that are substantially better than the general -O$x$ settings. BinTuner's outputs seriously undermine prominent binary diffing tools' comparisons. In addition, the detection rate of the IoT malware variants tuned by BinTuner falls by more than 50%. Our findings paint a cautionary tale for security analysts that attackers have a new way to mutate malware code cost-effectively, and the research community needs to step back to reassess optimization-resistance evaluations.

## 1 Introduction

Binary code, which is pervasive in our daily lives, spans a broad spectrum from traditional PC software, emerging IoT device firmware, to tremendous malware. As high-level language information such as data structures and types are missing in binary code, studying software security problems with only access to binary code is a challenging but also fascinating task [1–3]. Especially, the similarities between two binary code versions can reveal rich information even in the absence of source code. For example, whether a similar high-severity vulnerability recurs in other programs, or whether different malware variants belong to the same family. Therefore, binary diffing research generates a large body of literature on this topic, such as software vulnerability search [4–10], security patch analysis [11–13], malware similarity analysis [14–20], and code clone detection [21–25]. As pure syntax-based binary code representation (e.g., instruction mnemonic n-grams or data constants [26, 27]) are prone to false negatives, the trend of binary diffing technique is to overlook ostensible, syntactic differences and capture semantic similarities. At the other end of the spectrum, pure semantic similarity analysis is infeasible in practice due to its complexity and undecidability [28]. Existing approaches are more apt to adopt mixed syntactic/semantic code representations to measure binary code difference.

Compiler optimization is the most common factor leading to the semantics-preserving but syntactically different binary code. To achieve the goal of using less computing resources, modern compilers contain a large number of available optimization options, which can transform binary code notably [29]. For example, loop-related optimization (e.g., unswitching and loop unrolling) effectively rewrite control flow structure, and peephole optimization substitutes a loop-free code with an optimal assembly code sequence [30]. Therefore, evaluating the resilience against the changes caused by compiler optimization settings has become a convention for binary diffing tools. We surveyed the research literature in the last 12 years and assessed their resilience experiments.

*Corresponding author

We find that the impact of compiler transformation on binary code is limited by the *default* optimization levels. For example, Asm2Vec in IEEE S&P'19 [21] takes the comparison between O3 and O0 as the "most difficult" case. However, the optimization flags in GCC's -O3 setting only account for less than 48% of all available options. We argue that the power of compiler optimization on binary code difference has been significantly underestimated on modern CPU architectures.

In this paper, we first study compiler optimization effects on binary code differences. Then, we investigate the latent capability of all available optimization options on binary code differences. Our research is motivated by the usage of non-default optimization settings in practice. **First**, research papers [31–33] have confirmed that many performance-critical applications (e.g., programs running in resource-constrained devices) resort to a program-specific optimization sequence, which gains augmented improvements beyond default -Ox levels. **Second**, as quite a few binary diffing tools (e.g., MutantX-S [27], CoP [23], and BinSim [14]) work with adversaries, there is no reason to assume that software plagiarists or malware developers would restrict themselves to -Ox settings. For emerging IoT malware that has to run in miscellaneous embedded devices, traditional obfuscation techniques used in Windows malware (e.g., binary packing [34] and code virtualization [35]) are not well accepted because of the high runtime overhead and poor compatibility [36–38]. In contrast, compiling malware source code with different optimization flags other than the default levels can provide an additional layer of protection in metamorphism[1]. We have tracked the compiler provenance of Linux.Mirai family [40], an infamous IoT botnet, for one year. Surprisingly, we find that up to 42% of Linux.Mirai variants are not generated under default settings, and these variants reveal a much lower anti-malware detection rate than the rest of samples.

Our research methodology is inspired by search-based iterative compilation [41–45]. It has long been known that a fixed compiler optimization sequence does not produce optimal results in all cases. However, finding an optimal, program-specific compilation sequence is particularly challenging as well, because the search space of various optimization combinations is extremely large. Iterative compilation explores the huge optimization space using metaheuristic search algorithms. It attempts to find near-optimal or sufficiently good-enough solutions with acceptable overhead. The idea of iterative compilation is simple, but it can yield substantial performance gains. Therefore, the method of our study is to iteratively explore the optimization space to find better configurations than the default -Ox settings, so that the different degrees of binary code are greatly improved.

We developed an auto-tuning platform, named *BinTuner*, to maximize binary code difference. We apply the genetic algorithm to guiding optimization space exploration. The key step is to design a fitness function, which evaluates the results of compilation and steers the search process toward the optimal solutions. An efficient function can dramatically reduce the overall overhead of iterative compilation. We adopt normalized compression distance (NCD) as a simple-albeit-rudimentary fitness function to quantitatively measure binary code structural differences. NCD has a desirable theoretical underpinning in terms of Kolmogorov complexity [46], as well as superior performance.

We run BinTuner on SPEC integer benchmarks of CPU2006 and CPU2017, Coreutils, and OpenSSL. Our results show that at the cost of 279 to 1,881 compilation iterations, BinTuner can find various custom optimization sequences that outperform default settings in all 42 cases. For example, we obtain an additional improvement beyond LLVM's -O3 with an average value of 18% (peaking at 60%). Besides, we find that for Coreutils, the binary different degrees caused by -Os are greater than -O3 by 20%. The comparisons with prominent binary diffing tools show that their accuracies decline steeply for the tuned binary code produced by BinTuner. BinTuner's effect even surpasses Obfuscator-LLVM [47], a popular compiler-level code obfuscator at present. Our findings also reveal a new threat: cybercriminals can take a free ride of iterative compilation to automatically generate numerous metamorphic samples. We hope that our work spurs discussion and inspires the research community to redesign resilience evaluations for binary diffing approaches. In summary, our contributions are as follows.

- Binary diffing hinges on the comparison of mixed syntactic and semantic binary code representations, but the impact of compiler optimization on them is not well studied. Our work bridges this gap (§3).
- As far as we know, BinTuner is the first auto-tuning framework to deliver near-optimal binary code that maximizes the amount of binary code differences. Our findings highlight a pressing need for the research community to revisit the optimization-resistance experiments (§4 & §5).
- BinTuner can assist the binary diffing research in generating more diversified datasets for training and testing. Its source code and the tuned benign programs are available at (https://github.com/BinTuner/Dev).

The long version of this paper is available at (https://arxiv.org/abs/2103.12357).

## 2 Background & Motivation

For pedagogical reasons, we first characterize the flourishing binary diffing literature. Next, we discuss two representative binary diffing tools BinDiff and BinHunt. Our study treats BinHunt as an appropriate reference to evaluate BinTuner's outcome. At last, we introduce our observation on optimization-resistance experiments.

---

[1]Metamorphism means malware mutates code during propagations so that each variant exhibits little similarities to the other [39].

## 2.1 Binary Diffing Research

Even without access to source code, the similarities between two different binary code can expose the underlying relationship such as code clones, close malware lineage, or same toolchain provenance. Therefore, the multifaceted benefits of binary diffing have led to a wide adoption by various software security analysis tasks. Our long version in arXiv lists 47 top-venue papers related to binary diffing in the past 12 years. The selected papers cover the area of security, software engineering, programming languages, systems, and AI. These papers vary in the code representations to compare and how to measure their semantics similarities. The problems that they deal with include vulnerability search, malware analysis, patch inspection, plagiarism detection, and de-anonymizing code authors. In spite of these versatile applications, the accuracy of binary diffing is subject to modern compilers, which bring additional complexities to binary code structures [3]. The key to a binary diffing approach is to define a semantics-aware code representation, so that similar programs reveal the representations that are close to each other.

## 2.2 Mixed Syntactic and Semantic Binary Code Representations

The mixture of syntactic and semantic code representation strikes a balance between complexity and precision, and it is becoming a good practice. The lion's share of binary diffing papers we surveyed (42 out of 47) adopts the mixed syntactic/semantic representations. In particular, these methods differ in two levels: 1) which binary code structure is defined as code representation to compare (syntactic level); 2) how to represent and compare code representation semantics (semantic level). For syntactic level properties, most papers select recognizable binary code structures as code representations, including function, basic block, loop, trace, control flow graph (CFG), and call graph (CG). Their detection accuracies rest with *precisely locating the scope of such code representations*.

At the semantic level, the methods of measuring code representation semantics are even richer. They are ranging from computationally expensive but accurate to scalable but less robust properties. For example, symbolic execution represents the input-output relations as formulas and then verifies their equivalence using a theorem prover [14, 23, 48–50]; dynamic testing generates concrete inputs automatically to compare output values [7, 10, 51–53]; basic block re-optimization normalizes syntactically different data-flow slices to expedite scalable search [5, 54]; descriptive statistic features (e.g., the number of transfer instructions) gear towards fast matching target functions among large-scale binaries [8, 55]. Recent papers take advantage of deep learning and neural networks to learn the relationship between two binary code snippets [4, 21, 56–58]. For example, Asm2Vec [21] learns the lexical semantic relationships on x86/64 instruction set within a function scope, such as Streaming SIMD Extensions (SSE) operands are related to SSE registers, and file-related APIs are typically used together.

## 2.3 BinDiff & BinHunt

BinDiff [59, 60] is an industry-standard and the most-cited binary diffing tool. Many papers either rely on BinDiff's result or compare with BinDiff in their evaluations. BinDiff takes IDA's disassembly code [61] as input, and it relies on the comprehensive use of three-level statistic features (function, basic block, and the topological order of control flow/call graph) to achieve the goal of fast graph matching. BinDiff is resilient against moderate syntactic differences such as register swapping and instruction reordering.

BinHunt [62] is the first work to find semantic differences in binary code. We consider BinHunt as an improvement to BinDiff in two ways but at the cost of overhead. **First**, BinHunt applies symbolic execution and theorem proving to match functionally equivalent basic block pairs, so that it has a better resistance to intra-basic-block obfuscation types [14] than BinDiff. **Second**, BinHunt customizes a backtracking-style graph isomorphism algorithm to find the best matchings between functions and basic blocks. This algorithm can remove many false matches caused by BinDiff's graph matching heuristics. BinHunt's final difference score of two binary code varies from 0.0 to 1.0 (a higher score indicates more different). This score is a quantitative value to measure *the structural changes in CFG/CG and semantical changes in basic blocks*. Note that these changes are also the targets that BinTuner aims to achieve. Unfortunately, the computational cost of either BinHunt or BinDiff is too high to be acceptable as BinTuner's fitness function. In our evaluation, we treat BinHunt's score as an objective reference to verify whether BinTuner's outcome can beat -Ox levels. We present BinHunt score's calculation details in our long version.

## 2.4 Non-default Optimization Effects

All of 47 top-venue papers that we surveyed perform similarity analysis on the different versions produced by default compilation levels. Twenty-two papers claim they are resilient against the syntactical changes caused by compiler optimizations, and they treat the comparison between O3 (i.e., the highest general optimization level) and O0 (i.e., no optimization) as the worst case. However, both GCC's and LLVM's -O3 levels only contain less than 48% of the available compilation options. Security analysts have confirmed that compiler optimization effects can make malware analysis complicated. Qihoo 360's security analysts reported that the aggressive compiler optimization hinders the extraction of Mirai IoT botnet classification features [63].

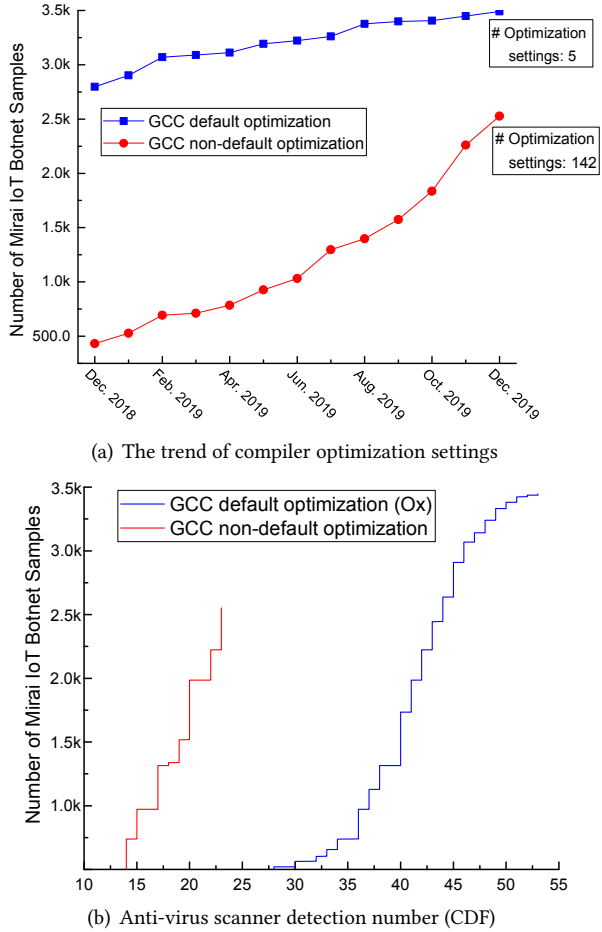

**Figure 1.** The data of Mirai IoT botnet family in 2019.

Since Dec. 2018, we have tracked the compiler provenance of Mirai IoT botnet family for one year. We leverage VirusTotal's Intelligence service [64] to collect the Mirai samples that have different hash values. Because Mirai's source code was leaked online in 2016 [65], we use BinTuner to generate a large training set with all applicable combinations of compiler versions and optimization levels, including non-default compilation options. We adopt BinComp's method [66] to reverse-engineer compiler provenance information (e.g., compiler family, compiler version, and optimization level) for each collected Mirai sample. Figure 1(a) shows that until Dec. 2019, up to 42% (2, 527) of Mirai variants are compiled by 142 kinds of GCC's non-default optimization settings. Note that different from PC malware, no Mirai samples apply packing or code virtualization. In spite of this, harnessing compiler optimization can still bypass the anti-virus detection. We confirm this by counting the recognition numbers of all available anti-virus scanners in VirusTotal. Figure 1(b) shows the cumulative distribution result of VirusTotal detection numbers. Obviously, the Mirai variants that are compiled by custom compiler optimization settings reveal a much better evasive effect than the rest of samples.

## 3 Compiler Optimization Effects on Binary Code Differences

In this section, we focus on the effects of compiler optimization on syntactic and semantic binary code representations. An in-depth understanding of these effects is crucial to the design of a robust binary diffing tool, but this problem is not well studied by the previous work.

### 3.1 Effect on Syntactic-Level Properties

Most binary diffing approaches assume the precise identification of code representation scopes before comparisons. Only in this way can they properly gauge their semantics. However, this assumption is fragile in practice. As binary function, basic block, and control flow graph are the three most common code representations (32 out of 47 papers that we surveyed compare them), we discuss how optimization algorithms can break the integrity of them.

#### 3.1.1 Function

Binary function scope is mainly affected by inter-procedural optimizations. The well-known function inlining optimization replaces function call instruction with the actual code of callee function. The frequently invoked library functions are most likely to be inlined. Although BinGo [7] proposes selective inlining to mitigate this problem, it is still quite ad hoc in the selection of function invocation patterns; Asm2Vec [21] adopts BinGo's approach to train its learning model, but it does not inline any library call; discovRE [8] even explicitly turns off function inlining in its vulnerable function search evaluation. Tail call optimization [67] is another obstacle to binary function recognition. Instead of using traditional call instruction, tail call switches to a jump instruction at the end of the caller function to target the callee function. This avoids the cost of frequent stack frame set-up and teardown. In the binary code of Coreutils compiled with GCC -O3, about 10% of functions use tail call optimization. Goër et al.'s work relies on dynamic instrumentation to recognize jump instructions as inter-procedural calls [68]. However, mainstream disassemblers and most binary diffing tools are all static-only approaches. As a result, tail call optimization will mislead their function matchings.

For random-sampling based function comparison methods [51, 52], they typically rely on calling conventions to recover complete function input parameters first. After that, they generate concrete values as function inputs and then compare function outputs. However, compiler optimizations may violate calling conventions and thus complicate function parameter extraction. For example, if the intended parameter value is already in an argument register, the compiler may not set that register explicitly at the function callsite.

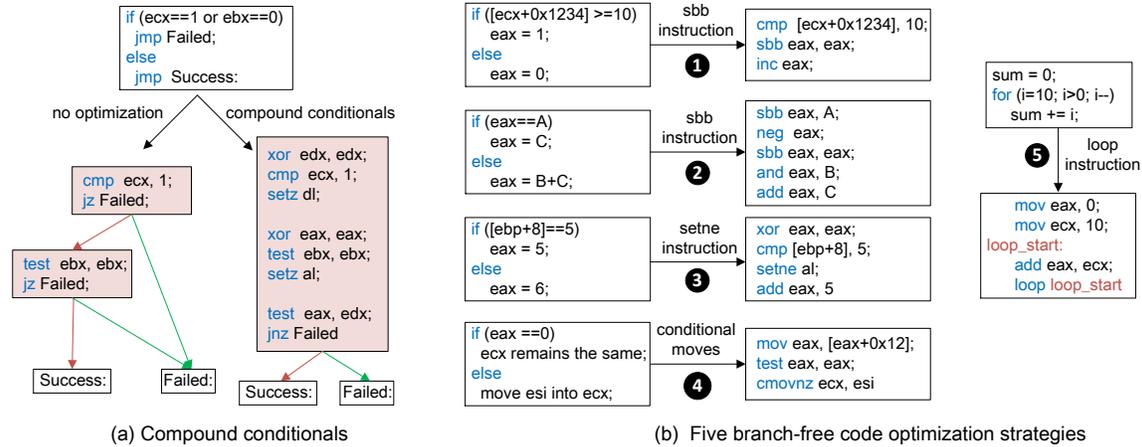

**Figure 2.** Compiler optimization breaks the integrity of basic blocks. (a) Compound conditionals generate more straight-line code by merging several basic blocks into one, and (b) five optimization strategies produce branch-free code to avoid a conditional jump instruction. All of them also change the structure of control flow graph.

Therefore, the absence of the value assignment instruction leads to the underestimation of function parameters.

### 3.1.2 Basic Block

Compared to the recovery of binary function's scope and parameters, the identification of basic block scope is much simpler. Expensive symbolic execution is typically performed within a basic block for accurate semantics modeling [10, 23, 48, 62]. The challenge here lies in that many intra-procedural optimizations (e.g., loop unrolling, compound conditionals, and basic-block merging) tend to produce branch-less code to favor pre-fetching instructions. As shown in Figure 2, modern compilers take advantage of the instruction side effect on FLAGS register (e.g., sbb, setz, and cmovnz) to avoid branches; while loop instruction does not set FLAGS at all, but it is exactly like dec ecx & jnz. Straight-line code can avoid branch misprediction and facilitate pipeline execution, but it also merges several basic blocks into one. Branch-free code violates the assumptions embodied by basic-block centric comparison models, because they are either straightforward "1-to-1" (one basic block in source function is matched against the one in target function) [10, 23, 48, 56, 62] or "n-to-n" [69]. Dealing with basic block merging requires heavyweight inter-basic-block control flow analysis.

### 3.1.3 Control Flow Graph (CFG)

A number of binary diffing methods measure CFG similarity [8, 10, 17] or match CFG structural features [55, 57]. However, CFG is more vulnerable to both inter-procedural and intra-procedural optimizations. Most factors that break the integrity of function and basic block (e.g., function inlining and loop-related optimizations) can effectively change the control flow graph structure as well. Another example is optimizing switch structure via binary search. Typically, the compiler translates a switch structure into an indirect jump and a lookup table for switch-case handlers, since it takes O(1) lookup time. The pattern looks like jmp dword ptr [eax*4 + Address]. Address is the lookup table starting address, and eax, controlled by switch's condition, is the index to a specific switch-case handler. However, if switch's cases are not in a small sequential range, both GCC and LLVM will adopt a binary search algorithm instead [70]. As a result, the new CFG will reveal more branches.

## 3.2 Effect on Semantic-Level Properties

Quite a few tasks [8, 16, 27, 55, 57, 71] need to perform a large-scale binary similarity analysis, such as malware clustering and bug search in firmware images. To meet the scalable goal, they represent the semantic as a vector of descriptive numeric features, such as the number of particular opcode types. However, the transformation of some compiler optimizations can generate totally different binary code snippets in syntax. This impedes the binary diffing work that does not extract the intrinsic semantics of code representations.

The arithmetic division is the most expensive integer calculation on CPU. Figure 3(a) shows the strength reduction optimization rewrites the division of a constant using multiplication [72]. The peephole optimization example in Figure 3(b) substitutes two instructions (10-byte length) with semantically equivalent but much faster instructions. The new ones only take as few as 3 bytes and have less fetch-execute cycles. Figure 3(c) shows an example of loop vectorization [73–75]: it makes use of modern CPU's fast SSE vector instructions to perform the same matrix product operation on multiple values simultaneously. To obtain the optimal performance, GCC has built-in implementations for many standard C library functions (e.g., strcpy and strcmp) [76]. Figure 3(d) optimizes the call to strcpy as a sequence of immediate-to-memory mov, which is much faster than its natural loop-based equivalent. Without semantics information, the lightweight, lexical-based features cannot find they are equivalent. The book "Hacker's Delight" [77] contains a collection of optimization tricks to speed up arithmetic algorithms via bitwise operations, and many of them have been accepted by LLVM and GCC as optimization options.

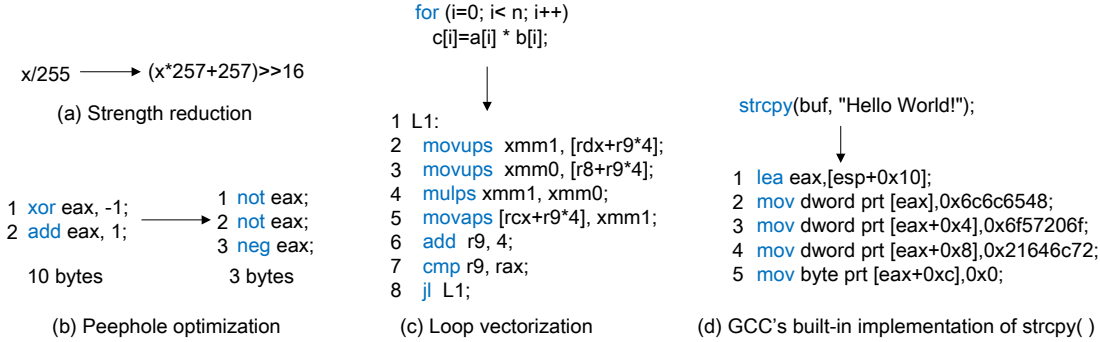

**Figure 3.** Compiler optimization generates a totally different binary code snippet in syntax. (a) x/255 is re-implemented via multiplication with a perfect approximation; (b) peephole optimization replaces non-optimized instruction segment with a faster set of instructions; (c) loop vectorization takes advantage of SSE vector instructions to run matrix product in parallel; (d) GCC's built-in implementation of strcpy() becomes a sequence of immediate-to-memory mov instructions.

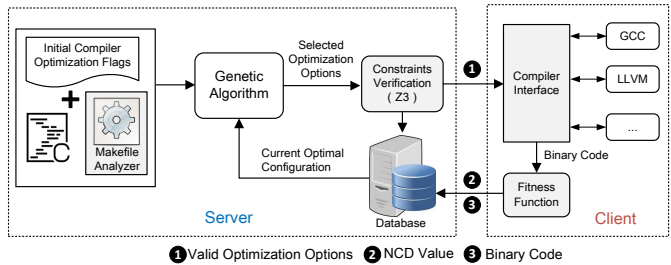

**Figure 4.** The overview of BinTuner's architecture.

## 4 BinTuner Design

We conduct an empirical study using iterative compilation to figure out *to what extent compiler optimization can change binary code*. Our study demonstrates that iterative compilation can automatically find much better optimization sequences, which work in concert to yield further improvements in binary code differences.

### 4.1 Overview

We build an auto-tuning framework, called *BinTuner*, to tune binary code differences via iterative compilation. Figure 4 shows the architecture of our framework. The core on the server side is a metaheuristic search (e.g., genetic algorithm) engine, which directs iterative compilation towards maximizing the effect of binary code differences. The client side runs different compilers and the calculation of the fitness function. Both sides communicate valid optimization options, fitness function scores, and compiled binaries to each other, and these data are stored in a database for future exploration. When BinTuner reaches a termination condition, we select the iterations showing the highest fitness function score and output the corresponding binary code as the final outcomes.

Similar to the observation of adaptive optimizing compiler [42], our rationale behind using genetic algorithm is that the options revealing the optimal effects on binary code difference are rare, but the local minima are frequent. In light of this, biased random search such as genetic algorithm can find good-enough solutions more quickly than local search such as hill climbing. We tune four parameters for the genetic algorithm, including mutation_rate, crossover_rate, must_mutate_count, and crossover_strength. As shown as Figure 4's grey boxes, BinTuner consists of four components.

**Makefile Analyzer.** BinTuner takes over the role of makefile to drive the multiple rounds of compilation and linking. We utilize the "scan-build" tool [78] to extract source file dependencies, configuration information, and initial compiler optimization flags from the target program's makefile.

**Constraints Verification.** Both LLVM and GCC explicitly specify a set of constraints between optimization flags, including adverse interactions and dependency relationships. In some cases, two options negatively influence each other, and turning on them together leads to a compilation error. Some other compilation options only work when a certain option has been activated. For example, -fpartial-inlining has any effect only when -finline-functions is turned on. To avoid compilation errors caused by such constraints, we manually translate them into logical first-order formulas offline after understanding the compiler manual. The knowledge we learned is easy to move between the same compiler series. We only need to consider the different optimization options introduced by the new version.

When BinTuner is running, constraints verification component uses a solver to check the correctness of newly generated optimization options. Otherwise, it will eliminate conflicting optimization sequences.

**Compiler Interface.** It works as a dispatcher loop to glue multiple compilers, genetic algorithm, and fitness function calculation. Compiler interface automates the whole iterative compilation process and is extensible for new compilers.

**Fitness Function.** Existing fitness functions do not suffice for BinTuner. We choose a new fitness function, Normalized Compression Distance (NCD), to quantitatively evaluate how close a given optimization sequence is to our expected optimum solution. The strategy for doing so is explored next.

### 4.2 Fitness Function Selection

A crucial step of genetic algorithm is to design a fitness function, which navigates the process of natural selection towards the optimal generations [79]. A qualified fitness function has to meet two requirements: 1) it can quantitatively determine how fit a solution is; 2) the calculation of fitness function should be efficient. Otherwise, it will become a performance bottleneck, and then the overall cost will increase drastically.

**Challenges.** Existing program-related fitness functions do not serve our need: quantitatively measuring the strength of binary code difference. We originally planned to use BinDiff or BinHunt difference score. Unfortunately, they do not satisfy the above second requirement: high efficiency. Their calculations have to disassemble the binary code first, which accumulates to significant overhead after multiple generations of genetic algorithm. In our evaluation, many benchmarks' binary code size are beyond 50M (up to 97M). Even on our powerful server machine, IDA [61] has to take 8 ~ 11 minutes to disassemble one large-size sample; BinDiff needs additional 5 ~ 9 minutes to complete comparison, and BinHunt's running time increases to 30 ~ 66 minutes. Therefore, we have to look for a low-computational-overhead measurement, which can approximate to the strength of binary code difference even without disassembly.

**Normalized Compression Distance.** The recent successes on large-scale malware classification using an information-theoretic measure, Normalized Compression Distance (NCD) [80–83], caught our attention. NCD infers the degree of similarity between arbitrary byte sequences by the amount of space saved after compression. Its theoretical merit comes from Kolmogorov complexity, which is algorithmic information theory that can measure code irregularity and randomness [46]. However, Kolmogorov complexity is uncomputable. Li et al. [84] proposed using a lossless data compression method (i.e., NCD) to approximate to Kolmogorov complexity. The calculation of NCD score is as follows.

$$NCD(x, y) = \frac{C(x \cdot y) - min(C(x), C(y))}{max(C(x), C(y))} \quad (1)$$

$C(x)$ represents a specific lossless compression algorithm, which returns the compressed length of program $x$'s code section in raw bytes; while $x \cdot y$ indicates the concatenation of two programs' code sections. NCD score ranges from 0.0 to 1.0 (the higher, the more different). If $x$ and $y$ are identical, the NCD score becomes 0.0. The accuracy of this approximation relies on the quality of compression algorithm, and recent malware classification work demonstrates that LZMA algorithm [85] is a good candidate [82].

**Correlation.** The intuition behind our selection of NCD is that the impact of compiler optimization on code representations causes structural irregularities in the binary code. Code regularity represents that certain code structures are repeated time after time. When compiling with O0 (i.e., no optimization), the compiler tends to generate boilerplate code. Various optimizations break the integrity of code structures, and hence the optimized code is more likely to exhibit irregularities. This explains that the binary code compiled under O0 setting typically has a much higher compression ratio than O3 version. In BinTuner's each iteration, we compute the NCD score between the existing solution and O0, and genetic algorithm prefers the optimization sequence that reveals a higher NCD score. We also calculated Pearson correlation values between NCD scores and BinHunt difference values for two relatively small SPEC benchmark programs: 462.libquantum & 429.mcf. The reason for selecting these two programs is that we can terminate BinHunt's experiments within a reasonable time. Our experimental results show that about 70% of significant positive correlations between NCD scores and BinHunt difference scores. We present detailed BinTuner's genetic algorithm and Pearson correlation value plot in our long version.

**NCD Calculation Performance.** The average NCD calculation time is less than 30 seconds. Taking NCD as the fitness function, BinTuner completes the above two experiments in *42 minutes*, while BinTuner takes 58.3/75.8 hours to terminate if we use BinDiff/BinHunt difference score as the fitness function. Using NCD as the fitness function speeds up BinTuner's performance by *two orders of magnitude*.

## 5 Evaluation

**Experimental Setup.** We use LZMA algorithm [85] in NCD calculation and Z3 [86] solver to remove conflicting optimization options. The testbed contains two Intel Xeon Gold 6134 processors and 256G memory, running Ubuntu 20.04 LTS. We terminate BinTuner's iterative compilation empirically when the successive NCD's growth rate is less than 0.35%. At this point, we treat the improvement of NCD is reaching the point of diminishing returns. Typically, we can obtain a set of best results that all reveal the same NCD score, and we select the last one to evaluate BinTuner's performance.

**Dataset.** We evaluate BinTuner with SPEC integer benchmarks including CPU2006 and CPU2017, Coreutils-8.30, and OpenSSL-1.1.1. In addition to measuring CPU performance, SPEC CPU2006 is often used as a complicated evaluation case for binary code analysis approaches in the past decade. SPEC CPU2017 is the latest generation of SPEC CPU benchmark with larger and more complex workloads. Compared to CPU2006, CPU2017 benchmarks have up to 2.3X more lines of source code and 10X higher dynamic instructions [87]. Our dataset contains the benchmark suites used for measuring CPU integer processing power: SPECint 2006 and SPECspeed 2017 Integer.[2] Coreutils and OpenSSL are the two most popular utilities in binary code search evaluations. Coreutils is a package of 95 utilities' executable code. In embedded systems, developers typically statically link them into one single binary code, so we do it in the same way in our evaluation. At last, we tune IoT malware to test their evasion capabilities to VirusTotal's anti-malware scanners.

### 5.1 BinTuner's Efficacy

The first experiment is to determine whether BinTuner can find custom compilation sequences that can cause additional enhancements in binary code differences. As most of the related work treats the comparison between O3 and O0 as the worst case, we also take O0's binary code as the baseline to calculate NCD during BinTuner's iterative compilation. We did not choose BinHunt's difference score as the fitness function due to its high computational overhead. Instead, we only compute BinHunt difference scores for several cases, including the binaries compiled by default -Ox settings and BinTuner's final outcomes, because we take them as objective references to verify whether BinTuner's results can outperform -Ox levels. BinHunt's comparison is well suited for us to interpret the compiler optimization impact on the structural changes in CFG/CG and semantical changes in basic blocks. Note that BinTuner's outputs retain functional correctness because all of BinTuner's outputs pass the test cases shipped with our dataset.

**LLVM.** Figure 5(a) shows BinHunt difference scores under multiple LLVM 11.0 optimization settings. We first look at the "O3 vs. O0" bar, which is taken by many binary diffing tools as the maximum difference in their compiler-agnostic evaluations. However, BinTuner's outputs, shown as the white bar, are better than "O3 vs. O0" in all cases with an average improvement of 18%. The peak value (as much as 60%) happens at 462.libquantum, in which BinTuner's result reveals large differences in the syntactic properties of code representations—only 19% of basic blocks and 27% of functions are matched with the -O0 version. Moreover, this benchmark involves quite a few factorizations of numbers and the dot product of matrix. These features enable strength reduction and loop vectorization to take optimization effects, and thus they also change semantic properties of code representations. When we only focus on default -Ox settings, we find that -O3 indeed works best for most cases, but its distance from -O2 is insignificant. Also, we notice two exceptions that -O1 (401.bzip2) and -O2 (625.x264_s) are slightly better than -O3. Besides, we also compare BinTuner's outputs with -O3 versions, and the last bar shows that they share small similarities in most cases.

**Table 1.** BinTuner's search iteration numbers and total running time (hour). The data of SPEC benchmarks are represented as (min, max, median).

|  | SPECint 2006 | SPECspeed 2017 | Coreutils | OpenSSL |
|---|---|---|---|---|
| LLVM |  |  |  |  |
| # Iterations | (347, 687, 470) | (279, 585, 415) | 527 | 593 |
| Hours | (0.3, 22.7, 0.8) | (0.3, 48.5, 0.6) | 4.4 | 4.9 |
| GCC |  |  |  |  |
| # Iterations | (491, 937, 612) | (469, 1881, 946) | 841 | 803 |
| Hours | (0.4, 31.3, 4.4) | (0.5, 70.9, 3.2) | 7.0 | 6.7 |

**GCC.** Similarly, Figure 5(b) presents the results under GCC 10.2 optimization settings. Compared with "O3 vs. O0", BinTuner's outputs obtain an average enhancement of 15%, and the tuned binary code of Coreutils achieves the maximum improvement of 55%. The custom compilation sequence completely messes up Coreutils's control flow graphs, in which BinHunt only matches 11% of graph edges with the O0 version, while "O3 vs. O0" has up to 37% matched CFG edges. Note that the average difference score of "BinTuner vs. O0" (0.77) is very close to the difference score of the wrong pair comparison of "Coreutils vs. OpenSSL" (0.79). This means the distance of "BinTuner vs. O0" is so significant that BinHunt is hard to distinguish it from the wrong-pair matches. Due to the space limit, the first bar of Figure 5(b) is "Os vs. O0". Os enables all of O2's optimizations except those increasing code size. However, somewhat counterintuitively, we observe an outlier: *Coreutils' GCC -Os presents an even larger amount of differences than GCC -O3 by 20%*. This is a counterexample of the long-held belief that O3 is always the best in the amount of binary code differences.

**Cross Comparison.** For the two most striking cases shown in Figure 5, We also present BinHunt's cross comparison results among BinTuner and -Ox levels in our long version. BinTuner's results are unquestionably the most significant ones.

**LLVM vs. GCC.** The vertical comparisons between Figure 5(a) and 5(b) reveals that the same benchmark may exhibit different patterns under different compilers. For example, BinTuner achieves the best improvement for 462.libquantum under LLVM. In contrast, GCC's default -Ox settings already work pretty well on the same program, leaving only marginal improvement space to BinTuner.

### 5.2 Compiler Optimization Impact on Code Similarity Representations

In this section, we zoom in on each bar of Figure 5 to present our findings behind quantitative difference values. The detailed metrics data are shown in our long version. We calculate the ratio of matched (basic blocks, CFG edges, and non-library function) under different compilation settings.

---
[2] We remove five benchmarks that have either compilation or linking errors: 403.gcc and 471.omnetpp for LLVM; 401.bzip2 and 464.h264ref for GCC, and 602.gcc_s for the both.

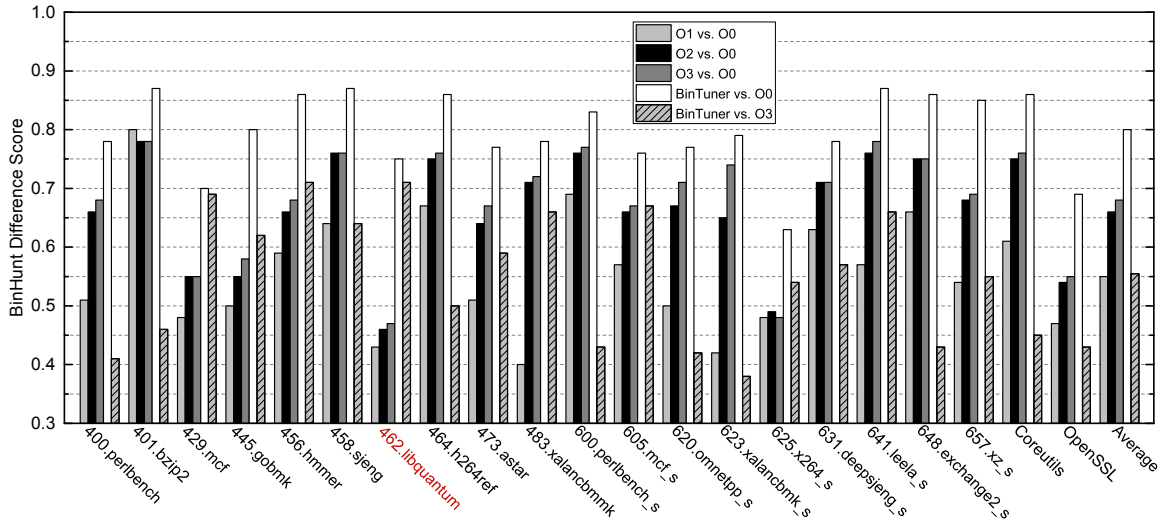

(a) LLVM 11.0: SPECint 2006, SPECspeed 2017, Coreutils, and OpenSSL

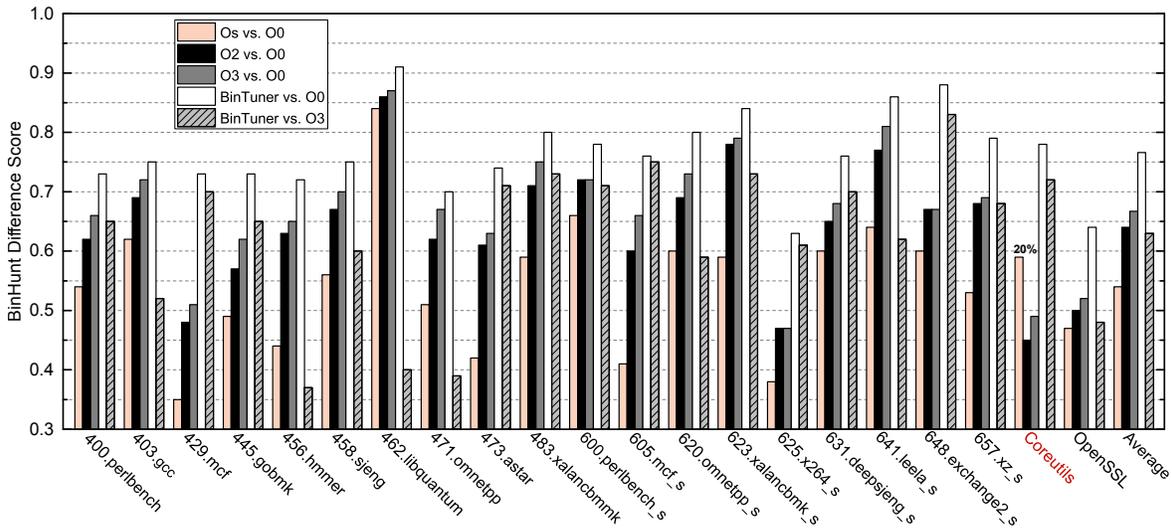

(b) GCC 10.2: SPECint 2006, SPECspeed 2017, Coreutils, and OpenSSL

**Figure 5.** BinHunt difference scores (the larger means more different) of our dataset under various optimization settings. "4\*\*" benchmarks belong to CPU2006, and "6\*\*" benchmarks are CPU2017. We highlight the most striking cases as red.

These data reflect the optimization impact on the most common code representations. In general, as the optimization level increases, the portion of matched code representations decreases. Especially, the binary compiled by -O3 does not represent the lower bound anymore, but BinTuner's outputs tend to produce more drastic changes. Among the three code representations, CFG is the most susceptible to compiler optimizations. For example, for 657.xz_s compiled by LLVM, the matched CFG edges drop sharply from 35% ("O1 vs. O0") to as little as 8% ("BinTuner vs. O0"). We also present the number of BinTuner's iterations and total running time, and we summarize them in Table 1. For 38 out of 42 tested programs, BinTuner reaches the termination condition within 1K iterations. BinTuner's performance bottleneck mainly comes from benchmark's compilation time, as long, one-time compilation time will accumulate to high cost after many iteration rounds. The worst case, 623.xalancbmk_s from SPECspeed 2017, has a pretty large code size and complicated library dependencies, and therefore both LLVM and GCC will take 6 ∼ 8 minutes to complete compilation and linking. Considering the extremely large search space, using NCD as the fitness function is cost-effective to find near-optimal compilation settings.

**NCD Variation.** We choose NCD as the fitness function, and Figure 6 plots NCD's variation over BinTuner's iterations for the four most significant test cases. Although each program shows different NCD patterns, the general trend is to steer genetic algorithm towards optimal solutions with

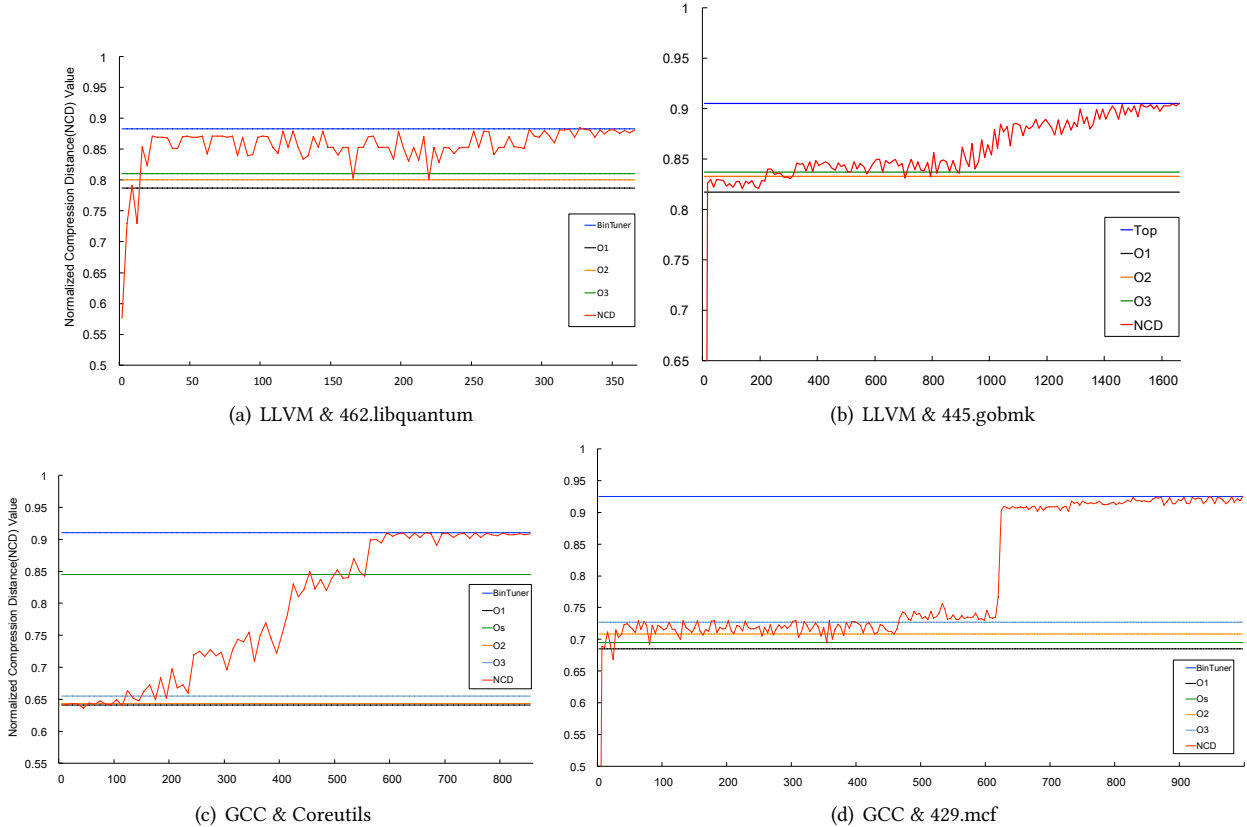

**Figure 6.** The aggregated NCD variation (higher is better) over BinTuner iterations. We show two most significant cases from LLVM and GCC, respectively. Note that the O1/O2 lines in (c) are very close and may be difficult to see.

| Flag | Potency |
| --- | --- |
| -funroll-loops | 18.0% |
| -fslp-vectorize | 13.2% |
| -fjump-tables | 12.6% |
| -finline-functions | 9.5% |
| -ftree-vectorize | 5.8% |
| -mlong-calls | 4.2% |
| -mstackrealign | 3.6% |
| -fwrapv | 3.2% |
| -fmerge-all-constants | 3.0% |
| -freg-struct-return | 2.2% |
| 94 other flags | 24.7% |

Jaccard Index (O3, BinTuner) = 0.54

(a) LLVM & 462.libquantum

| Flag | Potency |
| --- | --- |
| -funroll-loops | 17.8% |
| -fjump-tables | 12.5% |
| -fvectorize | 12.3% |
| -finline-functions | 10.6% |
| -ftree-vectorize | 4.8% |
| -mlong-calls | 4.5% |
| -fno-escaping-block-tail-calls | 4.1% |
| -fmerge-all-constants | 3.8% |
| -fwrapv | 3.6% |
| -fpcc-struct-return | 2.1% |
| 89 other flags | 23.9% |

Jaccard Index (O3, BinTuner) = 0.57

(b) LLVM & 445.gobmk

| Flag | Potency |
| --- | --- |
| -finline-small-functions | 9.4% |
| -ftree-vectorize | 7.9% |
| -freorder-functions | 7.3% |
| -funswitch-loops | 7.0% |
| -fpeel-loops | 6.9% |
| -fpeephole2 | 6.6% |
| -freorder-blocks | 6.2% |
| -ftree-loop-vectorize | 5.5% |
| -fbranch-count-reg | 5.1% |
| -falign-loops | 5.0% |
| 125 other flags | 33.1% |

Jaccard Index (O3, BinTuner) = 0.61

(c) GCC & Coreutils

| Flag | Potency |
| --- | --- |
| -finline-small-functions | 8.7% |
| -freorder-functions | 7.6% |
| -freorder-blocks-and-partition | 7.2% |
| -ftree-loop-distribute-patterns | 6.8% |
| -fpeephole2 | 6.7% |
| -ftree-vectorize | 6.4% |
| -fmove-loop-invariants | 6.0% |
| -floop-unroll-and-jam | 5.6% |
| -fbranch-count-reg | 4.7% |
| -falign-functions | 4.1% |
| 127 other flags | 36.2% |

Jaccard Index (O3, BinTuner) = 0.63

(d) GCC & 429.mcf

**Figure 7.** Top 10 most potent optimization flags for the significant benchmarks shown in Figure 6.

small fluctuations. Furthermore, when the termination criteria is met; that is, the improvement of NCD is reaching a plateau, *we can get multiple different versions that all reveal the best NCD score.* Recall that Coreutils' GCC -Os presents an even larger amount of differences than GCC -O3, so Os's NCD value is also larger than O3 in Figure 6(c). The NCD in Figure 6(d) jumps by 23% at the 620th iteration. We attribute this sudden leap to the mutation of genetic algorithm.

### 5.3 Optimization Flag Potency

For the significant cases shown in Figure 6, we try to understand which optimization flags contribute the most to the binary code differences obtained. This is not a trivial task, because figuring out the interactions among a set of optimization flags is challenging. To approximate the potency of each flag, given the optimal optimization sequence tuned by BinTuner, we measure the drop of BinHunt difference score when this flag is removed from that sequence. We normalize all BinHunt score drops to sum up to 100%. This measurement is not perfect, because some optimization flags may have competing or conflicting effects. Figure 7 presents the top 10 most potent optimization flags for the four significant benchmarks shown in Figure 6. We did not find such a standard "-Odiff" flag combination that can create a binary-different file, because each benchmark requires a different set of compilation options to achieve the best potency. However, our experiment still reveals interesting observations.

For the two significant benchmarks of LLVM (462.libquantum and 445.gobmk), they reach the optimal potency through a few large steps. They are dominated by the top four optimization options: loop unrolling, loop vectorization, switch-case optimization (-fjump-tables), and function inlining. Besides, 445.gobmk's top 10 flags contain tail call optimization, -fno-escaping-block-tail-calls, which can hide a binary function's boundary.

GCC's most potent optimization flag for Coreutils and 429.mcf is function inlining (-finline-small-functions), and other top flags take incremental steps with smaller potency effects. In addition to the flags that can change the CFG structure (e.g., loop-related optimizations), the top 10 flags in Figure 7(c) and (d) also reflect the compiler optimizations that break the integrity of basic block and affect semantic-level properties. For example, GCC's -freorder-blocks and -fbranch-count-reg favor producing branch-free code; peephole optimization (-fpeephole2) can mislead fast code matching approaches that compare numeric vector features.

Note that although most of these top 10 flags also appear in O3 sequence, the remaining flags are different from the ones in O3 sequence. At the bottom line of Figure 7, we show the share of common optimization flags between O3 and BinTuner's output using Jaccard index: $|A \cap B|/|A \cup B|$. Jaccard index results indicate that BinTuner can find different optimization options that yield further improvements in binary code differences.

### 5.4 Comparative Evaluation of Prominent Binary Diffing Approaches

We have demonstrated the efficacy of BinTuner by taking BinHunt [62] difference score as an objective reference. Given the same source code, BinTuner is able to generate drastically different binary code; this casts a doubt on whether the tuned binary code can also reliably complicate the analyses across multiple advanced binary diffing approaches. We conduct a separate experiment to compare prominent binary diffing tools. Unfortunately, only a very small portion of binary diffing papers release their source code.

**Tools' Selection.** We selected open-source binary diffing tools, including Asm2Vec [21], INNEREYE [56], VulSeeker [4], and BinSlayer [15]. In addition, we re-implement the method of CoP [23], IMF-SIM [51], and Multi-MH [10]. The superset of these seven tools and BinHunt is representative enough to cover the mixed syntactic/semantic binary code representations that we discussed in §3. Asm2Vec, INNEREYE, and VulSeeker are three machine learning based methods to learn the semantic similarities between two functions, basic blocks, and control flow graphs, respectively. IMF-SIM represents random-sampling based function comparison, which generates concrete inputs automatically to compare function outputs. CoP and Multi-MH are two examples of basic-block centric comparison. BinSlayer improves BinDiff [60] with the Hungarian algorithm for accurate graph matching. We did not test the dynamic approaches that compare system call dependency graph [18, 25] or aligned API call sequence [14], because they are difficult to measure the changes of binary code representations.

**Experiment Settings.** The challenge of comparing different binary diffing tools is that they adopt different similarity metrics such as graph edit distance [15] or statistical significance [48]; directly showing their similarity scores is not informative. We normalize their comparisons by calculating the ratio of truly matching function pairs that are also the rank #1 matching candidates (i.e., Precision@1). Precision@1 is also adopted by IMF-SIM [51] and Asm2Vec [21] to measure detection accuracy. We perform the comparative evaluation on Coreutils and OpenSSL, which are the most popular test suites in vulnerability search and code clone detection. We still take the O0 version as the baseline, and Figure 8 shows Precision@1 data under four different compilation settings. For the three machine learning based methods [4, 21, 56], we follow the same training data setting as Asm2Vec; that is, we train O0 functions to match the functions in other optimization settings. For the experiment of GCC & Coreutils, we test Os because Figure 5(b) shows the effect of "GCC -Os" on the binary differences of Coreutils is greater than -O3 by 20%. For the experiment of LLVM & OpenSSL, we also test Obfuscator-LLVM [47], a very popular compiler-level code obfuscator at present. When running Obfuscator-LLVM, we enable all three kinds of obfuscation schemes: instruction substitution, bogus control flow graph via opaque predicates, and control flow flattening.

**Results.** In summary, all tested binary diffing tools perform well when structural features are not changed too much. They show relatively high Precision@1 scores for the pair of O1 vs. O0. As the optimization level increases, interactions between multiple basic blocks become more intense, and structural properties are highly modified. Therefore, Precision@1 data fall into decline, and we can see a sharp drop when comparing with BinTuner's output. As both IMF-SIM and Asm2Vec papers also evaluate their tools using Precision@1, according to their worst cases, the values of Precision@1 caused by BinTuner are much lower than them by 46%~61%. These tested tools assume the integrity of function, basic block, or control flow graph. However, §5.2 has demonstrated that such an assumption is fragile. Among the three most common code representations, control flow structure is susceptible to a large number of optimization strategies. This explains why BinSlayer [15] starts a precipitous decline from O3 level, because it relies on bi-partite control flow graph matching.

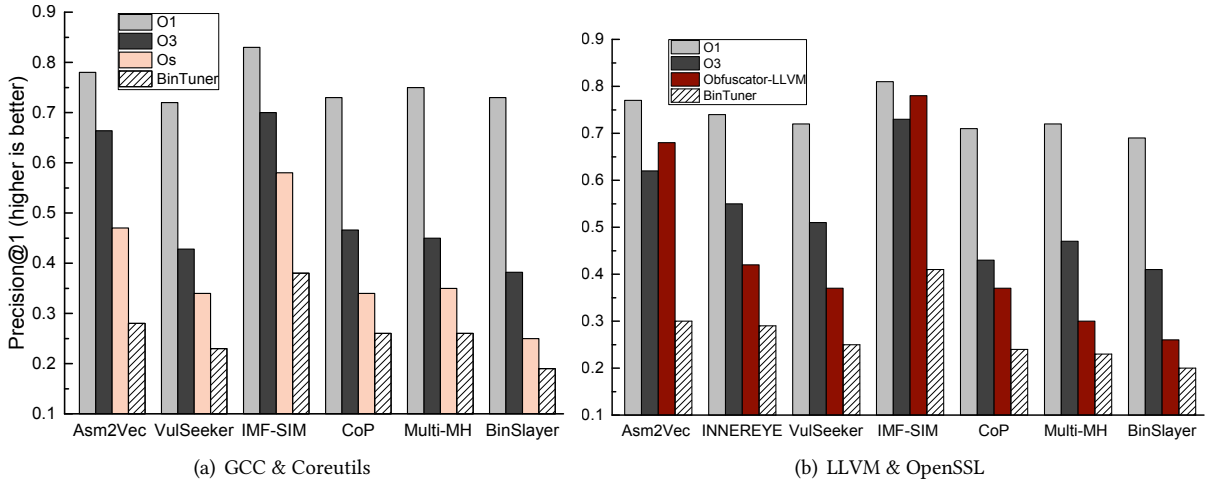

**Figure 8.** Precision@1 data (higher is better) reported by prominent binary diffing tools under four different settings. Note that 1) Os presents an even larger amount of differences than O3 in (a), and 2) INNEREYE [56] only works with LLVM.

**BinTuner vs. Obfuscator-LLVM.** Figure 8(b) shows that the potency of BinTuner is even better than Obfuscator-LLVM (O-LLVM). The reason is the obfuscation schemes applied by O-LLVM are limited in the function scope. By contrast, BinTuner has more options to achieve similar intra-procedural change effects. For example, O-LLVM's instruction substitution only contains several fixed rules to diversify arithmetic operations; while BinTuner enables peephole optimization [30], which has rich substitution rules to generate an optimal assembly code sequence. Furthermore, BinTuner contains inter-procedural optimizations to hide function call relationships (e.g., function inlining and tail call optimization). This viewpoint is also reflected in the experiment of IMF-SIM [51]. IMF-SIM outperforms the rest of tools. It treats the target functions as blackbox and performs dynamic testing to compare function-pair output values. Thus IMF-SIM is quite robust to intra-procedural optimization/obfuscation such as the effect caused by O-LLVM. However, BinTuner's custom optimization sequence leads to its loss of accuracy in two ways: breaking function's integrity and complicating the extraction of function parameters (see §3.1.1).

### 5.5 Tuning IoT Malware

The security risk motivating our research is that malware developers have utilized non-default compiler settings to generate metamorphic variants. Our one-year compiler provenance study on Mirai botnet presented in §2.4 confirms this new threat: 42% of them reveal different compiler optimization settings with -Ox. We apply BinTuner to the leaked source code of another two IoT botnet malware (LightAidra and BASHLIFE) [88] and count the anti-virus detection results via VirusTotal. Table 2 shows that the new malware variants tuned by BinTuner reveal different code features and bypass many anti-virus scanners. The detection number drops by more than half. Upon further investigation, the rest of anti-virus scanners can recognize the tuned samples because they match the signatures embedded in data section or API calls rather than code section. Like the trend shown in Figure 8, Asm2Vec and other tools also perform poorly against BinTuner generated malware samples. Table 2 demonstrates that, by taking advantage of iterative compilation, adversaries have a new alternative to evading detection.

**Table 2.** The number of anti-virus scanners recognizing IoT malware variants as malicious samples.

|                    | x86-32 | x86-64 | ARM | MIPS |
|--------------------|--------|--------|-----|------|
| LightAidra         |        |        |     |      |
| Default (GCC -O2)  | 46     | 42     | 44  | 43   |
| GCC -O3            | 45     | 41     | 43  | 41   |
| BinTuner           | 14     | 13     | 13  | 15   |
| BASHLIFE           |        |        |     |      |
| Default (GCC -O2)  | 41     | 37     | 39  | 38   |
| GCC -O3            | 40     | 37     | 38  | 37   |
| BinTuner           | 12     | 11     | 13  | 12   |

## 6 Related Work

We summarized binary diffing literature in §2. This section introduces the work most germane to BinTuner's design.

Our work differs from "Compiler-Generated Software Diversity" proposed by Jackson et al. [89] in a number of ways. Jackson et al.'s work aims to avoid that a single vulnerability compromises all vulnerable systems. Therefore, their diversification methods are designed to invalidate the hard-coded addresses of return-oriented programming (ROP) gadgets. However, they do not focus on changing CFG/CG structures or basic block semantics, and their diversified binaries still share many similarities that can be detected by BinHunt. In contrast, we take a free ride of iterative compilation to investigate to what extent compiler optimization can affect both syntactic and semantic binary code representations, which are the core in a binary diffing approach. To this end, we customize iterative compilation to favor adding structural differences to binary code.

Table 3. The average of execution speedup comparison.

|  | GCC | | LLVM | |
|---|---|---|---|---|
|  | O3 | BinTuner | O3 | BinTuner |
| SPECint 2006 | 6.6% | 4.7% | 7.1% | 5.0% |
| SPECspeed 2017 | 6.9% | 5.0% | 7.3% | 5.2% |
| Coreutils | 5.7% | 4.9% | 5.9% | 5.0% |
| OpenSSL | 5.9% | 5.8% | 6.0% | 7.2% |

Our study is inspired by Search-Based Software Engineering. Several papers share a similar idea to address software security problems. Closure* [90] looks for a sequence of JavaScript obfuscation schemes so that they can produce the optimal obfuscation potency. It also takes a guided stochastic algorithm to explore a huge search space. To quantitatively assess how difficult an adversary can understand an obfuscated JavaScript program, Closure* proposes an obscurity language model measuring code perplexity as the objective function. AMOEBA [91] iteratively performs a set of primitive code transformations to maximize the effect of software diversification. However, AMOEBA takes an empirical way to prune search space rather than using the metaheuristic search. This prevents AMOEBA from investigating the order in which code transformations can take more effect. Compared with the heavyweight code obfuscator such as Tigress [92], we view BinTuner's effect as a lightweight obfuscation strategy without adding noticeable computational overheads, but it still puts reverse engineers at a disadvantage.

## 7 Discussion and Future Work

We stress that our comparisons to binary diffing tools are not a criticism of their techniques, but rather offer a cautionary note for the evaluation of the compiler optimization resistance. We believe our study is inconclusive on this topic, but reporting our experiences will nevertheless raise awareness of compiler optimization on binary code differences. Please note that dynamic approaches that compare system call dependency graph [18, 25] or aligned API call sequence [14] are not affected by BinTuner.

The combination of iterative compilation and binary diffing shows promise, but BinTuner is still in its infancy. Although genetic algorithm is sufficient for producing diversified code, we plan to employ other advanced search heuristics (e.g., Markov chain Monte Carlo sampling [93]). Besides, utilizing the interactions between optimization options can further improve the search algorithm. For example, BinTuner explores all flags involving function inlining in proximity before moving to other groups.

Currently, we set up only one fitness function in BinTuner, so the tuned binary code may not present the best runtime performance. Table 3 shows the runtime speedup comparison, and we only find the execution speedup of OpenSSL caused by BinTuner can compete with O3. Next, like OpenTuner [45], we will study constructing custom optimization sequences that present the best tradeoffs between multiple objective functions (e.g., execution speed & NCD). To further reduce the total iterations of BinTuner, an exciting direction is to develop machine learning methods that correlate C language features with particular optimization options. In this way, we can predict program-specific optimization strategies that achieve the expected binary code differences.

## 8 Conclusion

Existing binary diffing's resilience evaluations are limited by the default optimization settings. In this work, we perform a systematic study using search-based iterative compilation. Our results demonstrate the effect of modern compiler optimization on binary code difference has been swept under the carpet for a long time. We wish our study can help the research community redesign the optimization-resistance experiments and evaluate the compiler-agnostic capability.

## Acknowledgments

We would like to thank our shepherd Yaniv David and the anonymous paper reviewers for their helpful feedback. We also thank VirusTotal for providing the academic API and malware samples. This research was supported by the National Science Foundation (NSF) under grant CNS-1850434.

The second student author, Michael Ho, passed away on February 12, 2018, after a courageous battle with leukemia. Michael was dedicated to this research project and made significant contributions. He was also a self-taught, talented magician and performed in many events. The audience always enjoyed his humor and creativity. We will remember his passion for research and life.

# Appendix
## A  BinHunt Difference Score Calculation

BinHunt's difference score of two binary code, ranging from 0.0 to 1.0 (a higher score indicates more different), is calculated via four steps:

1. Basic Block Matching Score: the matching score of two functionally equivalent basic blocks is assigned 1.0 if they use the same registers or 0.9 if they use different ones; otherwise, the score is 0.0.
2. Control Flow Graph Matching Score: ($\sum$ basic block matching scores) / (min(CFG1's size, CFG2's size));
3. Call Graph Matching Score: ($\sum$ CFG matching scores) / (min(CG1's size, CG2's size));
4. BinHunt's Difference Score: 1.0 - (CG's matching score).

## B  Metaheuristic Search: Genetic Algorithm

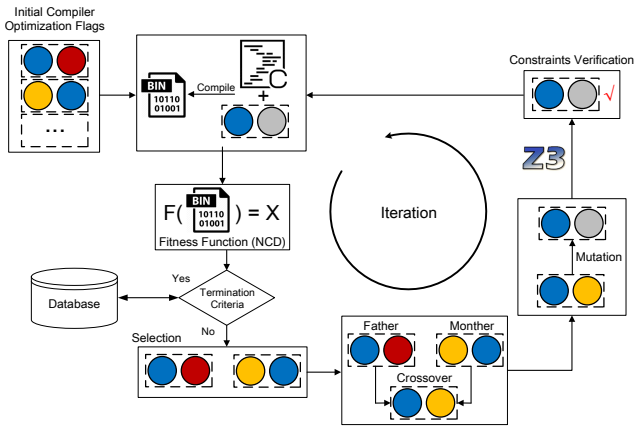

**Figure 9.** BinTuner's genetic algorithm iteration: the automatic selection of compiler optimization options towards maximizing the effect of binary code difference.

As illustrated in Figure 9, genetic algorithm (GA) encodes compiler optimization flags as a chromosome-like data structure, and then it applies crossover and mutation operations to these structures to simulate the evolution process under natural selection pressure. After the initial population and the selection of fitness function, pairs of parents go through the crossover to reproduce new individuals. These individuals will further mutate their genes to produce diversified populations. In addition to the traditional procedure, we add a constraint solver to verify the correctness of an optimization sequence and define a new fitness function to evaluate how individuals adapt to our needs. Following this style, this generational evolution is repeated until one of the following termination criteria is met.

- A parameterized number of iterations (e.g., 2K iterations).
- Allocated running time reached.
- The improvement of the fitness function score is reaching the point of diminishing returns.

## C  Correlation Between NCD and BinHunt

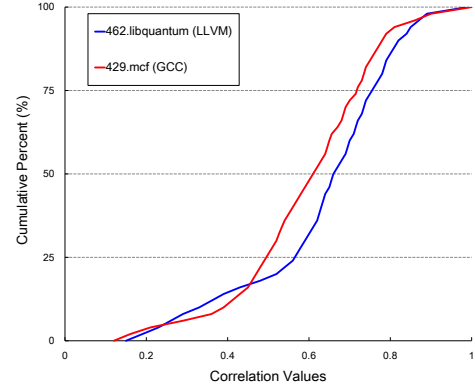

**Figure 10.** Pearson correlation values between NCD and BinHunt difference scores versus cumulative percent.

In statistics, when the Pearson correlation coefficient is greater than 0.4, it means there is a correlation between the two groups of data. A significant positive correlation happens when the coefficient is greater than 0.6. Figure 10 shows that the experimental results have about 70% significant positive correlations between NCD scores and BinHunt difference scores.

## D  Cross Comparison

For the two most striking cases shown in Figure 5, Table 4 and Table 5 further present BinHunt's cross comparison results among BinTuner and -Ox optimization levels. The last column shows the sum of cross comparison results for each compilation setting, and BinTuner's results are unquestionably the most significant ones.

**Table 4.** LLVM 11.0 & 462.libquantum

|          | O0   | O1   | O2   | O3   | BinTuner | Sum  |
|----------|------|------|------|------|----------|------|
| O0       | –    | 0.43 | 0.46 | 0.47 | 0.75     | 2.11 |
| O1       | 0.43 | –    | 0.17 | 0.21 | 0.73     | 1.54 |
| O2       | 0.46 | 0.17 | –    | 0.05 | 0.7      | 1.38 |
| O3       | 0.47 | 0.21 | 0.05 | –    | 0.71     | 1.44 |
| BinTuner | 0.75 | 0.73 | 0.7  | 0.71 | –        | 2.89 |

**Table 5.** GCC 10.2 & Coreutils

|          | O0   | O1   | Os   | O2   | O3   | BinTuner | Sum  |
|----------|------|------|------|------|------|----------|------|
| O0       | –    | 0.41 | 0.59 | 0.45 | 0.49 | 0.72     | 2.66 |
| O1       | 0.41 | –    | 0.49 | 0.22 | 0.29 | 0.71     | 2.12 |
| Os       | 0.59 | 0.49 | –    | 0.46 | 0.53 | 0.72     | 2.79 |
| O2       | 0.45 | 0.22 | 0.46 | –    | 0.15 | 0.73     | 2.01 |
| O3       | 0.49 | 0.29 | 0.53 | 0.15 | –    | 0.72     | 2.18 |
| BinTuner | 0.72 | 0.71 | 0.72 | 0.73 | 0.72 | –        | 3.6  |

**Table 6.** List of top-venue papers related to binary diffing in the last decade. All of them evaluate the binary code or bytecode produced by default compiler optimization settings. The fifth column summarizes the code representations that are used to measure similarity. "BB" and "CFG" are short for basic block and control flow graph, respectively. ● in the last column denotes this paper compares the proposed method with the industry-standard binary diffing tool: BinDiff; ◐ indicates this paper relies on BinDiff's output; and ○ means this paper mentions BinDiff as one of related tools.

| #  | Authors              | Conference   | Topic                                                                                  | Representation     | B.D. |
|----|----------------------|--------------|----------------------------------------------------------------------------------------|--------------------|------|
| 1  | Zhao et al. [94]     | CCS'20       | Memory corruption patch analysis by comparing input-related data structures.           | Memory objects     | ●    |
| 2  | Duan et al. [95]     | NDSS'20      | A unsupervised program-wide code representation learning for binary diffing.           | BB & CFG           | ●    |
| 3  | Yu et al. [96]       | AAAI'20      | A new model of neural networks that considers the order of CFG's nodes.                | BB & CFG           |      |
| 4  | Ding et al. [21]     | S&P'19       | Learning-based binary code clone search against compiler-level transformation.         | Function           | ○    |
| 5  | Zuo et al. [56]      | NDSS'19      | Cross-architecture basic-block similarity comparison using neural network.             | Basic block        |      |
| 6  | Gao et al. [97]      | FSE'18       | Learning-based vulnerability search via function semantic emulation.                   | Function           |      |
| 7  | Liu et al. [58]      | ASE'18       | Binary code similarity detection with deep neural network.                             | Function           | ●    |
| 8  | Gao et al. [4]       | ASE'18       | A semantic learning model that considers CFG and data flow graph (DFG).                | CFG & DFG          |      |
| 9  | David et al. [5]     | ASPLOS'18    | Scalable vulnerable procedure search in stripped firmware images.                      | BB slices          | ●    |
| 10 | Caliskan-Islam et al. [98] | NDSS'18 | De-anonymize the authors of binary code by detecting their unique coding style.     | Function           |      |
| 11 | Xu et al. [57]       | CCS'17       | Cross-platform binary diffing via neural network-based graph embedding.                | BB & CFG           | ○    |
| 12 | Wang et al. [51]     | ASE'17       | Identify similar functions in binary code using in-memory fuzzing.                     | Function           | ●    |
| 13 | Kargén et al. [99]   | ASE'17       | Scalable instruction trace alignment of binary code.                                   | Trace              | ○    |
| 14 | Ming et al. [14]     | Security'17  | Binary diffing via equivalence checking of API call sliced segments.                   | API call slices    | ●    |
| 15 | Xu et al. [12]       | ICSE'17      | A patch analysis framework to learn security patch/vulnerability patterns.             | BB & CFG           | ◐    |
| 16 | David et al. [54]    | PLDI'17      | Binary procedure similarity detection via re-optimizing basic block strands.           | BB slices          | ●    |
| 17 | Xu et al. [49]       | S&P'17       | Known cryptographic function detection via bit-precise symbolic loop mapping.          | Loop               | ●    |
| 18 | Chandramohan et al. [7] | FSE'16    | Scalable binary function matching via selective inlining.                              | BB tracelets       | ●    |
| 19 | Su et al. [100]      | FSE'16       | Code relatives detection via efficiently matching instruction dependency graph.        | Instruction graph  |      |
| 20 | Kalra et al. [101]   | FSE'16       | Detect application-affecting changes across two library binary versions.               | Function           | ●    |
| 21 | Feng et al. [55]     | CCS'16       | Scalable bug search in firmware images with CFG's numeric feature vector.              | BB & CFG           | ○    |
| 22 | Ding et al. [71]     | KDD'16       | Large-scale assembly function clone search using MapReduce subgraph search.            | Function           | ○    |
| 23 | David et al. [48]    | PLDI'16      | Compare similar binary procedures with small comparable basic block strands.           | BB slices          | ●    |
| 24 | Eschweiler et al. [8] | NDSS'16     | Cross-architecture search for similar binary functions using bipartite matching.       | BB & CFG           |      |
| 25 | Graziano et al. [102] | Security'15 | Compare malware code submitted to sandboxes to study malware developments.            | BB & CFG           | ◐    |
| 26 | Pewny et al. [10]    | S&P'15       | Cross-architecture bug search in binary code via basic block sampling.                 | BB & CFG           | ○    |
| 27 | Dalla Preda et al. [28] | POPL'15   | A formal semantic model for mixed syntactic/semantic binary similarity analysis.       | Basic block        | ◐    |
| 28 | Luo et al. [23]      | FSE'14       | Software plagiarism detection by matching equivalent basic block subsequence.          | Basic block        | ●    |
| 29 | Egele et al. [52]    | Security'14  | Search similar functions among binary code with dynamic similarity testing.            | Function           | ●    |
| 30 | David et al. [69]    | PLDI'14      | Binary function search by comparing fixed length of basic block tracelets.             | BB tracelets       | ○    |
| 31 | Sharma et al. [50]   | OOPSLA'13    | Verify the equivalence of two loops in binary code for loop optimizations.             | Loop               |      |
| 32 | Schuster et al. [103] | CCS'13      | Compare control flow traces of various inputs to identify backdoor regions.            | Trace              | ○    |
| 33 | Jang et al. [104]    | Security'13  | Recover the evolutionary relationship among a set of binary programs.                  | BB & Behavior      | ○    |
| 34 | Hu et al. [27]       | ATC'13       | Scalable malware clustering with instruction n-grams after generic unpacking.          | Instruction n-grams |     |
| 35 | Calvet et al. [53]   | CCS'12       | Cryptographic function detection in binary code with I/O parameter sampling.           | Loop               |      |
| 36 | Zhang et al. [105]   | ISSTA'12     | Algorithm plagiarism detection with dynamic value-based approaches.                    | Runtime invariants |      |
| 37 | Jang et al. [16]     | CCS'11       | Large-scale malware similarity detection using feature hashing.                        | N-gram & Behavior  | ○    |
| 38 | Chaki et al. [106]   | KDD'11       | Binary code provenance-similarity detection with supervised learning.                  | Function           | ○    |
| 39 | Jhi et al. [107]     | ICSE'11      | Software plagiarism detection by matching critical runtime invariant values.           | Runtime invariants |      |
| 40 | Rosenblum et al. [108] | ISSTA'11   | Recover the provenance of source language and compiler from binary code.               | Function           | ○    |
| 41 | Comparetti et al. [17] | S&P'10     | Identify similar malicious functionality that is not active at run time.               | BB & CFG           | ○    |
| 42 | Fredrikson et al. [18] | S&P'10     | Synthesize behavior-based malware specifications to match new malware.                 | Syscall graph      |      |
| 43 | Wang et al. [25]     | CCS'09       | Software plagiarism detection via system call dependence graph comparison.             | Syscall graph      |      |
| 44 | Hu et al. [19]       | CCS'09       | Find similar malware variants via fast function-call graph comparison.                 | Call graph         | ○    |
| 45 | Bayer et al. [20]    | NDSS'09      | Scalable malware clustering with behavioral profiles.                                  | Behavioral profile | ○    |
| 46 | Sæbjørnsen et al. [26] | ISSTA'09   | Binary clone detection by matching normalized instruction sequences.                   | Instruction n-grams | ○   |
| 47 | Brumley et al. [11]  | S&P'08       | Automatic 1-day exploit generation via patch difference analysis.                      | BB & CFG           | ◐    |

**Table 7.** Detailed comparison metrics under LLVM's optimization settings. The three numbers in each tuple mean the ratio of matched basic blocks, the ratio of matched CFG edges, and the ratio of matched non-library functions. The last two columns list BinTuner's search iterations and total running time (hour). "C.u." and "O.S." are short for Coreutils and OpenSSL, respectively.

| Prog. | O1 vs. O0 | O2 vs. O0 | O3 vs. O0 | BinTuner vs. O0 | # Iter. | Hr. |
|---|---|---|---|---|---|---|
| 400 | (29K/51K, 24K/75K, 1.9K/1.9K) | (27K/51K, 22K/79K, 1.7K/1.9K) | (26K/53K, 21K/84K,1.7K/1.9K) | (23K/63K, 18K/98K, 1.2K/1.9K) | 687 | 6.7 |
| 401 | (145/2K, 86/2.9K, 42/131) | (180/2.6K, 97/4K, 44/100) | (206/2.7k, 112/4.1k, 43/100) | (234/4K,140/6.6K, 49/101 ) | 519 | 0.5 |
| 429 | (309/410, 208/666, 50/52) | (312/525, 209/666, 50/52) | (313/525, 210/666, 50/52) | (329/560, 233/804, 50/58) | 482 | 0.4 |
| 445 | (18K/28K, 15K/38K, 2.7K/2.7K) | (16K/28K, 12K/38K, 2.5K/2.7K) | (16K/29K, 12K/43K, 2.5K/2.7K) | (13K/54K, 10K/84K, 2K/2.7K) | 577 | 3.4 |
| 456 | (6.5K/10K, 5K/14K, 609/612) | (5.8K/10K, 4.4K/15K, 541/609) | (6.5K/10K, 5K/14K, 609/612) | (1.6K/8.9K, 1.4K/13K, 145/612) | 436 | 1.1 |
| 458 | (2.5K/5.6K, 1.9K/8.1K, 185/190) | (2.3K/5.6K, 1.8K/8.1K, 171/185) | (2.5K/5.6K, 1.9K/8.1K, 185/190) | (1.7K/8.4K, 1.2K/14K, 88/185) | 347 | 0.5 |
| 462 | (878/1.2K, 673/1.5K, 154/155) | (838/1.2K, 661/1.6K, 135/154) | (853/1.3K, 661/1.7K, 135/154) | (232/1.2K, 194/1.5K, 42/154) | 363 | 0.3 |
| 464 | (9.4K/19K, 6.7K/25K, 643/643) | (8.1K/19K, 5.8K/25K, 571/643) | (7.2K/19K, 5.1K/26K, 571/643) | (4.8K/19K, 3.3K/25K, 285/643) | 457 | 3.2 |
| 473 | (904/1.4K, 600/1.7K, 181/192) | (879/1.4K, 597/1.9K, 118/192) | (856/1.4K, 583/2K, 118/192) | (351/1.4K, 962/1.7K, 62/192) | 371 | 0.3 |
| 483 | (67K/94K, 38K/91K, 28K/30K) | (45K/103K, 28K/137K, 12K/30K) | (44K/107K, 27K/143K, 12K/30K) | (57K/120K, 31K/119K, 10K/30K) | 584 | 22.7 |
| 600 | (48K/109K, 36K/167K, 3.1K/3.1K) | (45K/118K, 33K/183K, 2.4K/3.1K) | (44K/124K, 32K/195K, 2.4K/3.1K) | (44K/149K, 31K/230K, 1.6K/3.1K) | 549 | 13.8 |
| 605 | (468/835, 326/1K, 68/70) | (517/1.3K, 344/2K, 66/68) | (481/1.4K, 323/2.1K, 66/68) | (141/973, 88/1.4K, 38/68) | 316 | 0.3 |
| 620 | (28K/48K, 16K/51K, 11K/12K) | (21K/65K, 13K/87K, 6.3K/12K) | (21K/68K, 13K/94K, 6.2K/12K) | (20K/62K, 13K/83K, 6.2K/12.2K) | 585 | 28.7 |
| 623 | (74K/104K, 43K/104K, 29K/31K) | (49K/125K, 31K/171K, 12.8K/31K) | (48K/129K, 39K/178K,12.7K/31K) | (48K/125K, 30K/171K, 12.6K/31K) | 415 | 48.5 |
| 625 | (386/531, 308/667, 48/49) | (385/609, 285/839, 48/49) | (379/615, 292/848, 48/49) | (204/616, 126/876, 41/48) | 405 | 0.3 |
| 631 | (1.8K/3.8K, 1.4K/5.3K, 164/167) | (1.58K/3.8K, 1.21K/5.3K, 146/167) | (1.57K/3.8K, 1.19K/5.3K, 146/167) | (1.5K/4.3K, 1.1K/7K, 142/167) | 442 | 0.4 |
| 641 | (2.9K/5.8K, 1.8K/6.9K, 805/829) | (2.3K/6K, 1.5K/6.9K, 459/829) | (2.3K/6.7K, 1.4K/9.8K, 457/829) | (815/5.8K, 533/6.9K, 188/829) | 481 | 2.0 |
| 648 | (1.8K/2.9K, 1.5K/4.6K, 73/74) | (551/2.9K, 402/4.6K, 62/73) | (518/2.9K, 377/4.6K, 63/73) | (516/2.9K, 378/4.6K, 61/73) | 351 | 0.6 |
| 657 | (3.4K/5.6K, 2.6K/7.4K, 588/590) | (2.6K/5.8K, 2K/8.5K, 413/588) | (2.7K/6.4K, 2K/9.4K, 414/588) | (1.1K/6K, 759/9.2K, 170/588) | 279 | 0.6 |
| C.u. | (23K/44K, 16.7K/60K, 2.84K/2.85K) | (15K/44K, 10K/65K, 1.5K/2.85K) | (14.7K/44K, 9.7K/72K, 1.48K/2.85K) | (14K/44K, 11K/60K, 1.4K/2.85K) | 527 | 4.4 |
| O.S. | (10K/17K, 6.6K/22K, 2.1K/2.7K) | (8.9K/17K, 5.7K/22K, 1.7K/2.7K) | (8.8K/17K, 5.6K/22K, 1.68K/2.7K) | (8.6K/17K, 5.3K/22K, 1.66K/2.7K) | 593 | 4.9 |

**Table 8.** Detailed comparison metrics under GCC's optimization settings. The three numbers in each tuple mean the ratio of matched basic blocks, the ratio of matched CFG edges, and the ratio of matched non-library functions. The last two columns list BinTuner's search iterations and total running time (hour). "C.u." and "O.S." are short for Coreutils and OpenSSL, respectively.

| Prog. | Os vs. O0 | O2 vs. O0 | O3 vs. O0 | BinTuner vs. O0 | # Iter. | Hr. |
|---|---|---|---|---|---|---|
| 400 | (30K/49K, 28K/74K, 1.85K/2K) | (28K/49K, 23K/78K, 1.85K/2K) | (27.8K/64K, 22.2K/99K, 1.78K/2K) | (27K/81K, 21K/119K, 1.7K/2K) | 635 | 12.4 |
| 403 | (67K/171K, 59K/272K, 4.7K/5.7K) | (62K/172K, 48K/276K, 4.7K/5.7K) | (63K/219K, 48K/356K, 4.5K/5.7K) | (62K/205K, 49K/325K, 4.1K/5.7K) | 561 | 16.6 |
| 429 | (351/476, 305/608, 52/53) | (339/476, 305/608, 52/53) | (346/569, 232/812, 52/53) | (376/1.56K, 241/2.17K, 53/124) | 891 | 0.4 |
| 445 | (18K/28.9K, 15.7K/39.5K, 2.6K/2.74K) | (17K/28.9K, 13K/39.5K, 2.59K/2.74K) | (17K/38K, 12.4K/56K, 2.53K/2.74K) | (18K/56K, 13K/80K, 2.5K/2.74K) | 491 | 6.1 |
| 456 | (6.9K/10K, 7.1K/14K, 572/612) | (6.1K/11K, 4.7K/16K, 568/612) | (6K/13K, 4.5K/19K, 548/612) | (5.3K/11.8K, 4.1K/17K, 508/612) | 849 | 2.7 |
| 458 | (2.8K/5K, 2.6K/7.5K, 185/189) | (2.6K/5K, 2K/7.5K, 179/189) | (2.4K/5K, 2K/11K, 175/189) | (2.6K/8K, 2.3K/12.8K, 166/189) | 588 | 0.9 |
| 462 | (608/5K, 391/7.5K, 124/189) | (574/5K, 312/7.5K, 119/189) | (582/5K, 330/7.5K, 114/189) | (561/5K, 431/7.6K, 106/189) | 937 | 0.6 |
| 471 | (7.7K/13K, 5.9K/15K, 1.9K/2.68K) | (7K/14.7K, 4.5K/17.8K, 1.9K/2.68K) | (7K/12K, 4.3K/24K, 1.88K/26.8K) | (7.4K/21.5K, 4.9K/28K, 1.8K/2.68K) | 937 | 8.1 |
| 473 | (962/1.4K, 995/1.77K, 123/187) | (890/1.4K, 673/1.77K, 125/187) | (877/1.5K, 665/2.1K,121/187) | (1K/5.9K, 811/8.6K, 159/191) | 510 | 0.7 |
| 483 | (51K/96.8K, 39.6K/107K, 13K/29K) | (47K/138K, 30K/185K, 12.2K/29K) | (46K/179K, 29K/239K,12.1K/29K) | (46K/178K, 33K/249K, 12K/29K) | 573 | 31.3 |
| 600 | (49K/105K, 43K/154K, 2.76K/5.2K) | (46K/108K, 36K/165K, 2.66K/5.2K) | (46K/108K, 36K/165K, 2.66K/5.2K) | (47K/137K, 36K/202K, 2.48K/5.2K) | 946 | 23.8 |
| 605 | (520/757, 452/987, 68/71) | (471/778, 340/1K, 66/71) | (473/1.4K, 316/2.2K, 66/71) | (558/2.9K, 377/4.2K, 71/137) | 1,881 | 1.6 |
| 620 | (25K/46K, 19K/50K, 6.3K/11.5K) | (23K/60K, 15K/77K, 6.2K/11.5K) | (24K/87K, 15.7K/119K, 6.2K/11.5K) | (23.8K/88K, 15.7K/120K, 6.1K/11.5K) | 932 | 31.2 |
| 623 | (73.8K/148K, 58.9K/176K, 16K/30K) | (61.8K/294K, 40K/423K, 14.1K/30K) | (60K/328K, 37.9K/476K, 14K/30K) | (55K/366K, 31K/515K, 13.5K/30K) | 473 | 70.9 |
| 625 | (410/540, 391/648, 47/50) | (370/540, 286/695, 49/50) | (374/540, 278/710, 49/50) | (364/1.2K, 273/1.68K, 50/93) | 1,684 | 0.5 |
| 631 | (1.86K/3.6K, 1.6K/5.3K, 147/162) | (1.6K/3.7K, 1.24K/50K, 148/162) | (1.75K/4.9K, 1.3K/7.6K, 149/162) | (1.8K/5.8K, 1.39K/8.6K, 146/162) | 1,790 | 2.6 |
| 641 | (2.1K/4.9K, 1.6K/4.7K, 477/1.5K) | (1.9K/4.9K, 1.3K/9.8K, 426/1.5K) | (1.95K/11K, 1.2K/16K, 416/1.5K) | (2K/16.4K, 1.2K/23.6K, 435/1.5K) | 469 | 3.2 |
| 648 | (1.9K/3.5K, 1.6K/4.9K, 48/53) | (1.76K/3.5K, 1.36K/4.9K, 48/53) | (1.5K/3.5K, 1.2K/3.7K, 47/53) | (1.5K/8.1K, 1.1K/12K, 47/97) | 1,677 | 3.7 |
| 657 | (3.2K/5.6K, 2.98K/7.4K, 451/589) | (2.7K/5.8K, 2K/8.5K, 413/588) | (2.69K/6.4K, 2K/9.4K, 414/588) | (3K/11K, 2.1K/15.4K, 429/623) | 611 | 1.8 |
| C.u. | (19.5K/40K, 16.8K/56.2K, 1.8K/2.67K) | (23K/40K, 25.6K/59.5K, 1.69K/2.67K) | (21.8K/44.5K, 24.6K/66.8K, 1.5K/2.67K) | (18.5K/79.4K, 12.9K/114K, 1.4K/2.67K) | 841 | 7.0 |
| O.S. | (10K/15.8K, 7.5K/20K, 1.72K/2.71K) | (9.7K/15.8K, 6.7K/10K, 1.70K/2.71K) | (9.5K/15.8K, 6.5K/21.6K, 1.69K/2.71K) | (9.5K/19.7K, 6.4K/27.1K, 1.68K/2.74K) | 803 | 6.7 |